\documentstyle[preprint,aps]{revtex}\def\mathcal{\cal}

\input psfig

\begin{document}
\bibliographystyle{simpl1}
\date{August 26, 1996}

\title{Effect of Level Statistics on Superconductivity in
Ultrasmall Metallic Grains}

\author{Robert A. Smith$^{1, 2}$ and Vinay Ambegaokar$^2$} 

\address{$^1$School of Physics and Space Research, University of
Birmingham, Edgbaston, Birmingham~B15~2TT, United Kingdom}
\address{$^2$ Laboratory of Atomic and Solid State Physics,
Cornell University, Ithaca NY14853, USA}

\maketitle
\begin{abstract}
We examine the destruction of superconducting pairing in metallic grains
as their size is decreased for both even and odd numbers of electrons.
This occurs when the average level spacing $d$ is of the same order as
the BCS order parameter $\Delta$. The energy levels of these grains are
randomly distributed according to random matrix theory, and we must
work statistically. We find that the average value of the critical level
spacing is larger than that for the model of equally spaced levels for both
parities, and derive numerically the probability densities 
$\displaystyle P_{o,e}(d)$ that a grain of mean level spacing $d$ shows pairing.
\end{abstract}
\draft


A recent experiment by Black, Ralph and Tinkham \cite{BRT}
involving the observation
of a superconducting gap in ultrasmall Al grains (of size between
5 and 13nm) has led to reconsideration of an old but fundamental theoretical
question -- how small can a superconductor be? It is also of interest that
in a previous experiment \cite{RBT} on a smaller Al grain, 
the same group saw no sign
of a gap (although, as they noted, there are experimental difficulties in
observing an energy gap of similar magnitude to the average level spacing).
von Delft et al. \cite{DGTZ} have considered a simple mean field model for 
such a grain
which uses the standard BCS pairing interaction and assumes equal level
spacing for analytical simplicity. Even-odd parity effects \cite{JSA,GZ}, 
which can be seen in samples at least $10^4$ times larger \cite{LJEUDC,THTT}, 
and are of paramount importance
here, are included in their model. They find that the superconducting gap at
zero temperature should
cease at critical level-spacing $d_c^o=0.89\Delta(0)$ in odd grains, and
$d_c^e=4d_c^o$ in even grains, where $\Delta(0)$ is the zero-temperature bulk
gap. They also note that although the mean-field result is subject to several
types of correction, it does give a criterion for when pair correlations will
cease to exist. It is therefore surprising that in the data of BRT from
sample 4, an odd grain, that a gap is still seen although the sample is very
close to the odd critical level-spacing. We also note that BRT's data
shows no variation of the gap with level-spacing $d$, suggesting that their
samples are still on the flat part of the $\Delta(d)$ curve.

In this paper we consider the effect on the mean-field theory of relaxing the
condition of equal level-spacing. It is by now well-known that the
level-spacing in small metallic grains is the Wigner-Dyson (WD) distribution
\cite{WD}
obtained from random matrix theory (RMT) \cite{RMT}. 
This was first conjectured by
Gor'kov and Eliashberg \cite{GE}, and later proved by Efetov \cite{Efe}. 
The reason for
considering this effect is that most of the other corrections to mean-field
theory seem to lead to a reduction in $d_c$; on the other hand, level statistics
effects lead to larger values of $\langle d_c\rangle$, as we shall see. 

The first thing we shall do is to reproduce the results 
of von Delft et al.
\cite{DGTZ} for $d_c^o$ and $d_c^e$. We do this to demonstrate how the positioning of
the energy levels enters into the calculation, and how this leads to the
factor of 4 in the result $d_c^e=4d_c^o$. Our starting point is the
mean-field self-consistency equation
\begin{eqnarray}
\label{MFSC}
\frac{1}{\lambda}=d\sum_{|i|<\omega_c/d}\frac{1}{2E_i}
\left(1-2f_i\right)
\end{eqnarray}
where $E_i=\sqrt{(\epsilon_i-\mu)^2+\Delta^2}$, with $\epsilon_i$ the $i$-th
energy level, $\mu$ the chemical potential, $\omega_c$ the Debye energy, 
$d$ the level spacing,
and $\lambda$ the BCS interaction. The occupation factor $f_i$ differs
for even or odd parity ensembles
\begin{eqnarray}
\label{Focc}
f_i=\frac{f_i^+Z_+\pm f_i^-Z_-}{Z_+\pm Z_-}
\end{eqnarray}
where $f_i^{\pm}=\pm(e^{\beta E_i}\pm 1)^{-1}$ and 
$Z_{\pm}=\displaystyle\prod (1\pm e^{-\beta E_i})$. 
We will work at zero temperature,
so that $f_i=1/2$ if the chemical potential lies on a level, and zero otherwise.
In the case of equal level
spacing the chemical potential lies half-way between the last filled
and first empty levels in the even case, and  on the half-filled level
in the odd case, as shown in Fig. (1). 
For the case of the critical level spacing, the solution
has $\Delta(T=0)=0$, so that one has
\begin{eqnarray}
\label{Delc}
\frac{1}{\lambda}=\sum_{i=1}^{\omega_c/d_c^e} \frac{1}{i+1/2}
\quad;\quad
\frac{1}{\lambda}=\sum_{i=1}^{\omega_c/d_c^0} \frac{1}{i}.
\end{eqnarray}
These can be rewritten in terms of the digamma function to yield
\begin{eqnarray}
\label{Delc1}
\begin{array}{rclrcl}
\displaystyle\frac{1}{\lambda}&\!\!=\!\!&\psi(\omega_c/d_c^e)-\psi(1/2)\approx
\log{(\omega_c/d_c^e)}-\psi(1/2)\\
\displaystyle\frac{1}{\lambda}&\!\!=\!\!&\psi(\omega_c/d_c^0)-\psi(1)\approx
\log{(\omega_c/d_c^o)}-\psi(1).
\end{array}
\end{eqnarray}
Finally, since we know that $\psi(1)=-\gamma$, $\psi(1/2)=-\gamma-2\ln{2}$,
where $\gamma$ is the Euler-Mascheroni constant, it follows that
\begin{eqnarray}
\label{Delcres}
d_c^o=\frac{1}{4}e^{\gamma}\omega_c e^{-1/\lambda}=
\frac{1}{2}e^{\gamma}\Delta(0)\approx 0.89 \Delta(0)~;\quad d_c^e=4d_c^o.
\end{eqnarray}
We see that the factor of $4$ between $d_c^e$ and $d_c^o$ comes from the
fact that $\psi(1)-\psi(1/2)=2\ln{2}$, and thus ultimately from the
positioning of the chemical potential relative to the energy levels.
Furthermore if we write this out as a series for $2\ln{2}$,
\begin{eqnarray}
\label{logfour}
\begin{array}{rcl}
2\ln{2}&\!\!=\!\!&\psi(1)-\psi(1/2)\\[12pt]
&\!\!=\!\!&\displaystyle\left(\frac{2}{1}-\frac{2}{2}\right)+
\left(\frac{2}{3}-\frac{2}{4}\right)+\left(\frac{2}{5}-\frac{2}{6}\right)
+\dots
\end{array}
\end{eqnarray}
we see that only a few terms are needed before we are close to the final
answer. In other words, the value of this particular factor is
determined by a few energy levels near to the Fermi surface.

Let us now see how we can extend the above approach to the case of
non-equally spaced energy levels. Consider first the odd case, so that
the chemical potential lies on the half-filled level $\epsilon_0$.
The gap equation can then be written as
\begin{eqnarray}
\label{MFSC1}
\begin{array}{rcl}
\displaystyle\frac{1}{\lambda}&\!\!=\!\!&d\displaystyle\sum_i\frac{1}
{2\sqrt{(\epsilon_i-\epsilon_0)^2+\Delta^2}}\\
&\!\!=\!\!&\displaystyle\int_{-\omega_c}^{\omega_c}
\frac{d\omega}{2\sqrt{\omega^2+\Delta^2}}
d\sum_i \delta(\omega-\epsilon_i+\epsilon_0).
\end{array}
\end{eqnarray}
We can then take the average of this equation over the disorder ensemble,
so that all the statistical information about the level spacing occurs
in the average over the sum of delta functions. This can be related to the
two-level correlation function (TLCF) of the system as follows. The TLCF
is defined by
\begin{eqnarray}
\label{TLCF}
R(\epsilon-\epsilon')=d^2\langle\sum_{i,j} \delta(\epsilon-\epsilon_i)
\delta(\epsilon'-\epsilon_j)\rangle.
\end{eqnarray}
Since this is a function only of the energy difference,
$\epsilon-\epsilon'$, we may set $\epsilon'=0$ to obtain the result
\begin{eqnarray}
\label{Reqn}
\begin{array}{rcl}
R(\epsilon)&\!\!=\!\!&d^2\langle\displaystyle\sum_{i,j}\delta(\epsilon-\epsilon_i)
\delta(\epsilon_j)\rangle\\
&\!\!=\!\!&d^2\langle\displaystyle\sum_{i,j}\delta(\epsilon-\epsilon_i+\epsilon_j)
\delta(\epsilon_j)\rangle\\
&\!\!\approx\!\!&d\langle\displaystyle
\sum_{i,j}\delta(\epsilon-\epsilon_i+\epsilon_j)\rangle.
\end{array}
\end{eqnarray}
The averaged odd-gap equation can finally be written as
\begin{eqnarray}
\label{SCodd}
\frac{1}{\lambda}=\int_0^{\pi\omega_c/d} 
\frac{dx}{\sqrt{x^2+(\pi\Delta/d)^2}}R(x)
\end{eqnarray}
where $x=\pi\omega/d$. Since the system has time-reversal invariance, the 
TLCF given by RMT is that for the orthogonal ensemble \cite{RMT},
\begin{eqnarray}
\label{Rort}
R(x)=1-\frac{\sin^2{x}}{x^2}-\frac{d}{dx}\left(\frac{\sin{x}}{x}\right)
\int_x^{\infty}dt \frac{\sin{t}}{t}
\end{eqnarray}
and the average critical level spacing $\langle d_c^o\rangle$ is then
the solution of
\begin{eqnarray}
\label{dcodd}
\frac{1}{\lambda}=\int_0^{\pi\omega_c/d} \frac{R(x)}{x}.
\end{eqnarray}
This integral can then be performed analytically to give
\begin{eqnarray}
\label{dcoddres}
\langle d_c^o\rangle=\pi e^{\gamma+\pi^2/16-7/4}\Delta(0)=1.80\Delta(0)
\end{eqnarray}
and we see that the average odd critical level spacing is a factor $2$
larger than in equal level spacing model.

Next let us consider the even case. The chemical potential is now halfway
between the last filled level, $\epsilon_0$, and the first filled level,
$\epsilon_1$. It follows that the gap equation can now be written in the
form
\begin{eqnarray}
\label{SCeven}
\frac{1}{\lambda}=d\sum_i 
\frac{1}{2\sqrt{(\epsilon_i-\frac{\epsilon_0+\epsilon_1}{2})^2+\Delta^2}}
\end{eqnarray}
where the sum over $i$ includes both $\epsilon_0$ and $\epsilon_1$. For
$\epsilon_i$ not equal to $\epsilon_0$ or $\epsilon_1$, we can rewrite
\begin{eqnarray}
\label{levspac}
\epsilon_i-\frac{\epsilon_0+\epsilon_1}{2}=(\epsilon_i-\epsilon_1)
+\frac{\epsilon_1+\epsilon_0}{2}.
\end{eqnarray}
We know from the odd case that the distribution of the 
$\epsilon_i-\epsilon_1$ is described by the TLCF, $R(x)$. The distribution
of the $\epsilon_1-\epsilon_0$ is given by the nearest level spacing
distribution, $P(y)$. There is no analytic expression for $P(y)$, but it
is well approximated by the ``Wigner surmise'' \cite{RMT},
\begin{eqnarray}
\label{Wig}
P(y)=\frac{y}{2\pi}e^{-y^2/4\pi}
\end{eqnarray}
where $y=\pi(\epsilon_1-\epsilon_0)/d$. Let us now assume that the
distributions of $\epsilon_i-\epsilon_1$ and $\epsilon_1-\epsilon_0$ may
be treated independently. The actual distribution function we need is a
three-level function for $\epsilon_0$, $\epsilon_1$, and $\epsilon_i$, but
such a function is not discussed in the RMT literature.
The equation for the critical level
spacing, $d_c^e$, is then
\begin{eqnarray}
\label{dceven}
\frac{1}{\lambda}=\int_0^{\infty} dy P(y)\frac{2\pi}{y}+
\int_0^{\pi\omega_c/d_c^e} dx \int_0^{\infty} dy \frac{R(x)P(y)}{x+y/2}
\end{eqnarray}
where the first term comes from the levels $\epsilon_0$ and $\epsilon_1$,
which have to be treated separately, and the second term comes from all
other levels. Note that if we were to replace the denominator $x+y/2$ by
$x$ in the second term we should recover the odd integral. It follows that
we should evaluate the difference between the second term and the odd
integral, from which we obtain the result
\begin{eqnarray}
\label{dcevenres}
\langle d_c^e\rangle =\exp{(\pi-2I/\pi)}\langle d_c^0\rangle
\end{eqnarray}
where $I$ is the integral
\begin{eqnarray}
\label{Idef}
\begin{array}{c}
I\!\!=\!\!\displaystyle\int_0^{\infty}dt~t^2 e^{-t^2/\pi}\displaystyle
\int_0^{\infty}\frac{1}{x(x+t)}
\times\\[12pt]
\displaystyle\left[1-\frac{\sin^2{x}}{x^2}+si(x)\frac{d}{dx}
\left(\frac{\sin{x}}{x}\right)\right].
\end{array}
\end{eqnarray}
This integral cannot be performed analytically, and has the numerical
value $I\approx 1.7343$. We can therefore summarize the results for the
mean critical spacings in terms of the bulk BCS gap $\Delta(0)$ or the critical
spacing for equidistant levels in odd grains $d_c^o$ by
\begin{eqnarray}
\label{dcres}
\langle d_c^o\rangle=1.80\Delta(0)\approx 2.0 d_c^o ~,\quad
\langle d_c^e\rangle=7.67\langle d_c^o\rangle\approx 15.5 d_c^o.
\end{eqnarray}

We see that the consideration of level statistics not only makes both
the odd and even critical level spacings larger, it also increases the
ratio between them. The reason for this is that both the individual gap
equations, and the difference between the gap equations, involve the
inverse of energy level spacings. The fluctuations to smaller level
spacings are thus weighted more than those to larger than average level
spacings i.e. $\langle1/\delta E\rangle > 1/\langle\delta E\rangle$. 

In the analytic discussions above we have evaluated the mean
value of the critical level spacing. We note that the mean is only one
statistical measure of a probability distribution, and may not actually be
the one we want. We would therefore like to look at the probability 
distributions $P_{o,e}(d)$ of there being a superconducting gap in odd/even
grains with average level spacing $d$. We might imagine an experiment in
which many grains of the same nominal size are produced and examined for
the presence of a superconducting gap; the experimental results would
then yield $P_{e,o}(d)$. To obtain $P_{e,o}(d)$ we proceed numerically,
obtaining sets of energy levels $\{\epsilon_i\}$ by diagonalizing $N\times N$ random
matrices. Since the eigenvalues produced have a semicircular density of 
states \cite{RMT},
\begin{eqnarray}
\label{semi}
\rho(\epsilon)=\frac{1}{\pi}\sqrt{2N-\epsilon^2}~\theta(2N-\epsilon^2)
\end{eqnarray}
where $\theta(x)$ is the Heaviside function, we use the rescaling
\begin{eqnarray}
\epsilon\rightarrow\frac{1}{2\pi}\left[2N\sin^{-1}{(\epsilon/\sqrt{2N})}
+\epsilon\sqrt{2N-\epsilon^2}\right]
\end{eqnarray}
to obtain eigenvalues with average spacing unity \cite{MWL}. 
From the gap equations we see that the criterion
for a grain to have a non-zero superconducting gap is (all energies now in units of
$d$)
\begin{eqnarray}
\label{defsc}
\frac{1}{\lambda} < \sum_{|\epsilon_i-\mu|<\omega_c/d}
\frac{1}{2|\epsilon_i-\mu|}
\end{eqnarray}
where $\mu=\epsilon_0$ in the odd case, and $\mu=(\epsilon_0+\epsilon_1)/2$
in the even case. As in the analytical calculations, 
$d$ enters only in the upper cut-off $\omega_c/d$.
We choose the value $\lambda=0.193$ corresponding
to $\omega_c/d_c^o=100$ in the equal level spacing case, and $d$ is measured
in units of this $d_c^o$. For $d$-points from 0.5 to 1.0 we use 1000
realizations of $500\times 500$ matrices; for $d$-points from 1.0 to 4.0 
we use 1000 realizations of $300\times 300$ matrices; and for $d$-points from
4.0 to 20.0 we use 10000 realizations of $100\times 100$ matrices. We note
that for larger $d$, where we need more realizations to get good statistics,
we are fortunate in that we require smaller matrices. The results are shown in
Fig. (2). 

We see that for both odd and even cases there is significant chance
of superconductivity persisting beyond the critical level spacings deduced from
the equal level spacing model. Both curves also show long tails which we believe
are due to the non-zero probability of finding two levels very close together.
These long tails make it hard to estimate the mean value of the critical
spacing from the numerical data---though it is worth noting that Fig. (2) is
quite consistent with the analytic results in Eq. (20).  The long tails
also imply
that the mean is perhaps not the best measure for a typical critical spacing.
If instead we the value of $d$ where $P(d)=0.5$, we get about $1.7d_c^o$
for the odd case, and $6.5d_c^o$ for the even case. So, using this measure
we recover a factor of order $4$ between the two critical spacings.

Let us now discuss the implications of the above calculation for
experiment. First let us ask the question of what the BRT experiment
actually measures for the cases of even and odd grains. In both these
cases, since an electron tunnels onto and then off the island, the result
involves some sort of comparison between odd and even states. So do we see
the odd gap, even gap, or some mixture thereof? We will always work in the
zero temperature limit, which is effectively where the experiment is
performed. Consider the case of the
even grain. The lowest state an electron can tunnel into is the first
unoccupied level, so the energy cost should be the energy difference
between ground states of the system with $2N$ and $2N+1$ electrons. From
the $T=0$ limit of Eq. (4.9) of Ref. \cite{JSA}, this is given by
\begin{eqnarray}
\label{evenf}
E(2N+1)-E(2N)=\mu+\Delta.
\end{eqnarray}
To see which $\Delta$ is involved, note that $\Delta$ arose from
formulas (2.19) of Ref. \cite{JSA} which give
\begin{eqnarray}
\label{omdif}
\begin{array}{rcl}
\Omega_o-\Omega_e &\!\!=\!\!& \displaystyle
\frac{1}{\beta}\ln{\left(\frac{Z_{+}+Z_{-}}{Z_{+}-Z_{-}}\right)}\\[12pt]
&\!\!=\!\!& \displaystyle\frac{1}{\beta}\ln{\left(
\frac{1+(1-2N_{eff}e^{-\beta\Delta_e})}
{1-(1-2N_{eff}e^{-\beta\Delta_o})}\right)}.
\end{array}
\end{eqnarray}
At zero temperature we see that it is the odd gap, $\Delta_o$, 
that is measured. 

For the
case of an odd grain, the lowest state for the electron to go into is
the singly occupied state, and we measure the energy difference
\begin{eqnarray}
\label{oddf}
E(2N+2)-E(2N+1)=\mu-\Delta_o.
\end{eqnarray}
It follows that the lower branch of the odd grain curve should be the mirror
image of the even grain curve, so that $\Delta_o$ appears in
experiments on both even and odd grains.  We note that the first excited
state of an odd grain is
obtained by putting the extra electron into the first unoccupied state,
thus giving a state with two unpaired electrons. This should have an
energy roughly $2\Delta$ above the ground state, but the evaluation 
is complicated by the quasi-particles reducing
the phase space for pairing correlations.\cite{JVD}  

\begin{acknowledgements}
We would like to thank J. von Delft, I.~V. Lerner, M.~W. Long 
and D.~C. Ralph for useful discussions and comments. Support from
EPSRC Grant GR/J35238 (R.A.S.) and NSF Grant DMR-9121654 (V.A.) is
gratefully acknowledged.  V.A.\ thanks L. P. Gor'kov for a stimulating
comment, and the Department of Physics at the University of Florida
(where some of this work was begun) for hospitality.

\end{acknowledgements}

\vfil\eject

\begin{figure}
\centerline{\psfig{file=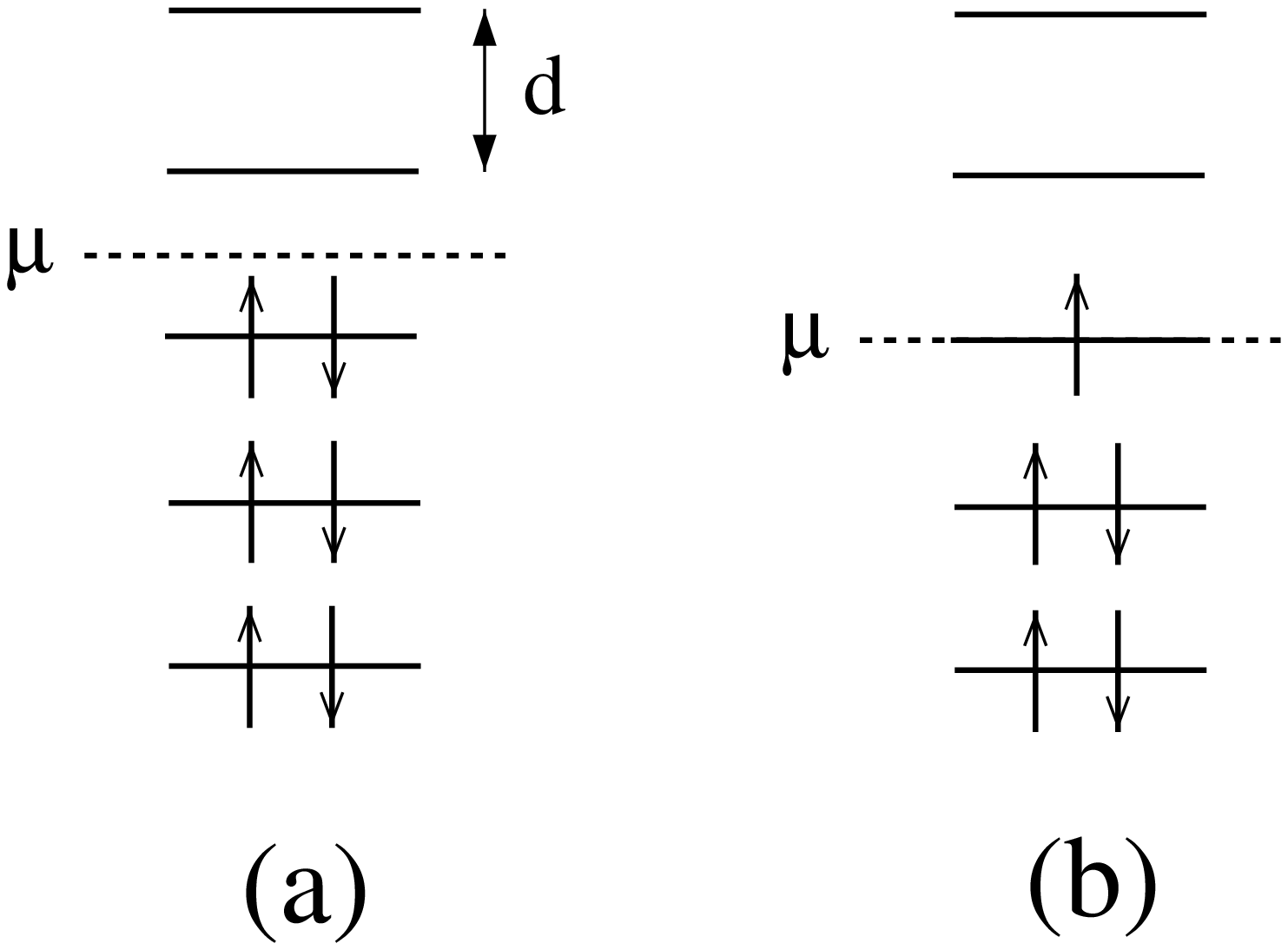,width=6truein}}
\vspace{0.5truein}
\caption{
The positioning of the chemical potential relative to the electronic
energy levels in a superconducting grain with (a) an even number of electrons;
(b) an odd number of electrons. In the even case the chemical potential lies
halfway between the last filled and first empty level; in the odd case it lies
on the half-filled level. Although illustrated for the equal level spacing
case, the same occurs for randomly spaced levels.}
\end{figure}

\newpage

\begin{figure}
\centerline{\psfig{file=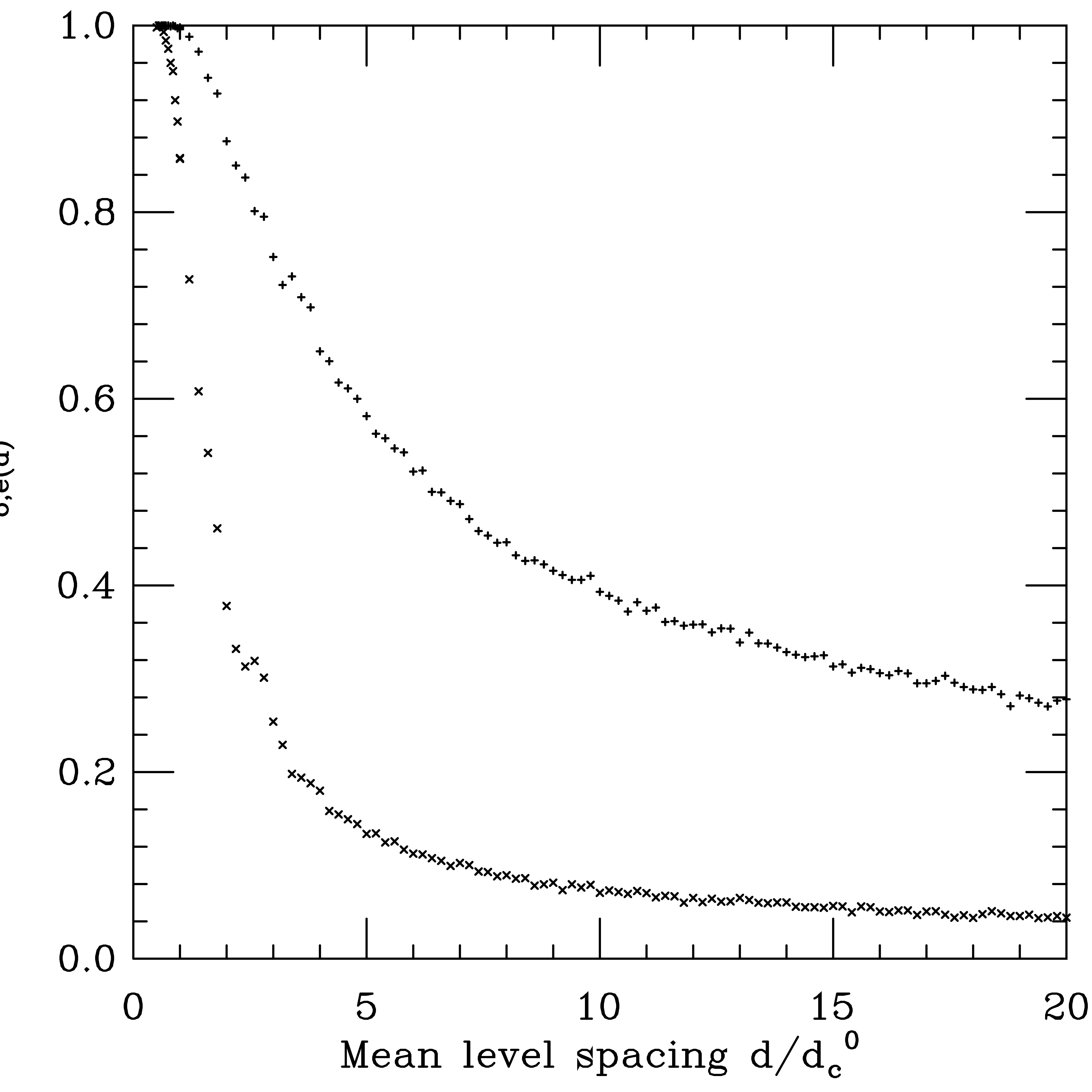,width=6truein}}
\vspace{0.5truein}
\caption{
The probability densities $P_{e,o}(d)$ of a metallic grain with 
mean level spacing $d$, and an even ($+$ symbols) or odd($\times$ symbols)
 number of electrons, having a non-zero 
superconducting energy gap. $d$ is measured relative to $d_c^0=0.89\Delta(0)$,
the critical level spacing for the equidistant model with an odd
number of electrons}.
\end{figure}


\begin{references}

\bibitem{BRT} 
C.~T. Black, D.~C. Ralph, and M.~Tinkham, Phys.\ Rev.\ Lett.\
{\bf 76}, 688 (1996).

\bibitem{RBT}
D.~C. Ralph, C.~T. Black, and M.~Tinkham, Phys.\ Rev.\ Lett.\
{\bf 74}, 3241 (1995).

\bibitem{DGTZ}
J.~von Delft, D.~S. Golubev, W.~Tichy, and A.~D. Zaikin, 1996 preprint.
 
\bibitem{JSA}
B.~Jank\'o, A.~Smith, and V.~Ambegaokar, Phys.\ Rev.\ B {\bf 50}, 1152
(1994).

\bibitem{GZ}
D.~S. Golubev and A.~D. Zaikin, Phys.\ Lett.\ A {\bf 195}, 380 (1994).
 
\bibitem{LJEUDC}
P.~Lafarge, P.~Joyez, D.~Esteve, C,~Urbina, and M.~H. Devoret, Phys.\
Rev.\ Lett.\ {\bf 70}, 994 (1992).

\bibitem{THTT}
M.~T. Tuominen, J.~M. Hergenrother, T.~S. Tighe, and M.~Tinkam, Phys.\
Rev.\ Lett.\ {\bf 69}, 1997 (1992).

\bibitem{WD}
E.~P. Wigner, Ann. Math. {\bf 53}, 36 (1991); F.~J. Dyson, J. Math. Phys.
{\bf 3}, 140 (1962).

\bibitem{RMT}
M.~L. Mehta, {\it Random Matrices} (Academic Press, Boston, 1991).

\bibitem{GE}
L.~P. Gor'kov and G.~M. Eliashberg, Zh. Eksp. Theor. Fiz. {\bf 48},
1407 (1965) [Sov. Phys. JETP {\bf 21}, 940 (1965)].

\bibitem{Efe} 
K.~B. Efetov, Adv.\ Phys. {\bf 32}, 53 (1983).

\bibitem{MWL}
We thank M.~W. Long for pointing this out and consequently reducing the
computing time needed.

\bibitem{JVD}
J. von Delft et al., to be published
\end{references}
\end{document}